\documentstyle[aps]{revtex}

\def\r{\mbox{\bf r}} 
\def\k{\mbox{\bf k}}

\begin{document}
\input epsf
\draft
\twocolumn[\hsize\textwidth\columnwidth\hsize\csname
@twocolumnfalse\endcsname
\preprint{PURD-TH-99-03, CERN-TH/99-21}
\date{February 1999}
\title{Quantum Dew}

\author{S. Khlebnikov$^{a}$
%\footnote{email address: skhleb@physics.purdue.edu}  
and I. Tkachev$^{a,b,c}$
%\footnote{ email address: Igor.Tkachev@cern.ch}
}
\address{
$^a$Department of Physics, Purdue University, West Lafayette, IN 47907, USA\\
$^b$TH Division, CERN, CH-1211 Geneva 23, Switzerland \\
$^c$Institute for Nuclear Research, Russian Academy of Sciences,\\
60th October Anniversary Prosp. 7a, Moscow 117312, Russia \\
}

\maketitle

\begin{abstract}
We consider phase separation in nonequilibrium Bose gas with
an attractive interaction between the particles. Using numerical integrations on 
a lattice, we show that the system evolves into a state that
contains drops of Bose-Einstein condensate suspended in uncondensed gas.
When the initial gas is sufficiently rarefied, the rate of formation
of this quantum dew scales with the initial density as expected for a process
governed by two-particle collisions.
\end{abstract}
\pacs{PACS: 98.80.Cq, 03.75.Fi\hspace{1.0cm} PURD-TH-99-03,~~~CERN-TH/99-21
\hspace{1.0cm} hep-ph/9902272}

\vskip2pc]

%\newpage
%\narrowtext
Theory of interacting Bose gases has been an important part of 
quantum statistical mechanics ever since Bogoliubov's seminal work
\cite{Bogoliubov}. In particular, 
{\em nonequilibrium} Bose gases are interesting from
at least two points of view. First, such gases can now be produced in
laboratory via modern cooling techniques.
A dramatic demonstration of the resulting
nonlinear dynamics is Bose-Einstein condensation (BEC) 
observed in alkali vapors \cite{exp1,exp2,exp3}. Second, nonequilibrium 
gases of elementary particles
frequently arise in cosmological scenarios and could have played an 
important role in the evolution  of the universe.

One way how nonequilibrium Bose gases arise in cosmology is via 
decay of a coherently oscillating field. This 
mechanism could be important, for instance, at the end of
an inflationary stage, i.e. during the reheating after inflation.
Indeed, it has been found that in some inflationary models,
the oscillating
inflaton field decays rapidly and completely into a 
gas that contains 
both the inflaton quanta and other types of Bose particles \cite{decay}.
These gases have very large occupation numbers 
in low-momentum modes and almost no occupation in high-momentum
modes; they are highly nonthermal.
The possibility of existence of such gases is by no means 
limited to the postinflationary era. In particular, there are indications
that nonbaryonic cold dark matter
constitutes a significant fraction of the matter in the universe at present. 
At the epoch of galaxy formation, gravitational instability develops on 
a variety of scales, which may lead to formation of small-scale dark 
matter clumps in galaxy halos. Dark matter particles trapped 
in a gravitational well are out of thermal equilibrium and are 
nonrelativistic. 
Typically interparticle interactions are very small, 
but, if the particles are bosons, the relaxation time can in certain cases
be comparable to the age of the universe \cite{it91}. 
This opens a possibility of
Bose-Einstein condensation and formation of Bose-stars \cite{rb,it86,csw}.
One proposed precursor of those is axion miniclusters \cite{kt_bs},
but modern particle models contain a variety of other fields
of potential interest in this respect: majoron, dilaton, moduli, to name
a few.

Thus, it is important to investigate the evolution of 
nonequilibrium Bose gases under various conditions. If the interaction between 
the particles is {\em repulsive}, and the energy density is sufficiently
low, a  Bose-Einstein condensate will form.
The process of Bose-Einstein condensation in this case has been studied
theoretically in a number of papers \cite{it91},\cite{BEC,ST,yale}.

The question we want to address in this paper is what happens to 
a nonequilibrium Bose gas if the interaction between its particles is
{\em attractive}, at least within a certain range of interparticle 
distances. There is hardly any doubt that an attractive 
interaction will lead to clumping and phase separation, and statements
to that effect have appeared in recent literature \cite{neg}.
However, it has remained unclear whether the clumps will
be in the normal or the superfluid state. In addition, kinetics of 
the clumping needs to be elucidated. Our main
result, obtained via numerical integrations, is that the clumps are
drops of Bose-Einstein condensate, i.e. each of them is characterized by a 
macroscopic order parameter. These drops remained suspended in 
uncondensed gas for as long as we could follow the evolution, although
they did grow somewhat at the expense of the gas. 
Because Bose condensation in the drops is attributable to the quantum 
statistics of the particles, we call such drops {\em quantum dew}.

The coherent, macroscopically ordered nature of quantum dew may be
important in cosmological (as well as laboratory) applications. 
Suppose for example that the particles it is 
made of can decay into some other particles. The macroscopically
populated mode of a coherent clump may work as a laser \cite{it86,it87}; 
as a result, quantum dew may decay much faster than an incoherent clump 
would. 

The purpose of the present paper is to prove the coherent nature of the 
clumps and to study the kinetics of appearance and growth of dew drops.
For this purpose, we have chosen the simplest model with an attractive 
interaction and, nevertheless, a stable ground state. From the point of
view of cosmological applications, perhaps the most important effects left out
of this simple model are the expansion of the universe and the gravitational
attraction. The expansion dilutes gas available for clumping and thus
slows the clumping down. The gravitational attractions works in the opposite direction.
The net effect of these 
opposing tendencies can in principle be found via numerical 
integrations, and we plan to return to this important question in future. 

The model contains a nonrelativistic complex Bose field
$\psi$ with the following equation of motion
\begin{equation}
2m i \frac{\partial \psi}{\partial t} = -  \nabla^2 \psi - 
|\psi|^2 \psi + g_{6} |\psi|^4 \psi \; .
\label{eqm}
\end{equation}
The field is normalized so that the attractive cubic term on the 
right-hand side has the coefficient of unity. The corresponding 
coupling $g_4$ then appears in the commutation relation:
\begin{equation}
[a_{\k}, a^{\dagger}_{\k'}] = |g_4| \delta_{\k\k'} \; ,
\label{com}
\end{equation}
the annihilation operators $a_{\k}$ being defined
via $\psi(\r) = V^{-1/2}\sum_{\k} a_{\k} \exp(i\k\r)$ in a finite volume $V$.
The quintic term in (\ref{eqm})
is repulsive, and it becomes important when
$\psi^{\dagger} \psi$ approaches $g_6^{-1}$.
(The coupling $g_4$ appearing in (\ref{com}) is related to $\lambda$ of
the relativistic $\lambda \phi^4/4$ potential via $g_4=3\lambda/2m$,
and to the scattering length $a$ of a nonrelativistic Bose gas via
$g_4 = 8\pi a$; here $\hbar=1$. The physical density is
$\psi^{\dagger} \psi /|g_{4}|$.)

Our integrations are set up as follows. 
In the initial state the 
occupation numbers $n_{\k} =a^{\dagger}_{\k } a_{\k}$ have a Gaussian
distribution over momenta
\begin{equation}
n_{\k} = A \exp(-k^{2}/k_{0}^{2})  \; .
\label{nk}
\end{equation}
The population of the homogeneous mode is $n_0=A$, which is not considerably
larger than population of other modes with small $k$. In this sense, 
there is no macroscopic condensate in the initial state.

Now, we assume that $g_4$ in (\ref{com}) is small compared to $A$ in
(\ref{nk}). Then, we can neglect the commutator of $a$ and $a^{\dagger}$
compared to the typical magnitude of $a$ itself, cf. Bogoliubov 
\cite{Bogoliubov}. As a result, the problem becomes classical and can 
be integrated on a lattice. This classical approximation 
has been used to study nonlinear dynamics of relativistic Bose fields at large occupation numbers \cite{class}
and the process of Bose-Einstein condensation \cite{yale}. In the present 
work, we use it to demonstrate the formation of quantum dew in the model 
(\ref{eqm}). This use involves no contradiction of terms: quantum dew
is an effect of quantum statistics when one thinks in terms of
individual particles, but it comes out as an effect of classical 
evolution in the  collective, field-theoretical description. Notice 
that eq. (\ref{nk}) determines only absolute values of $a_{\k}$; their 
phases are chosen as uncorrelated random numbers.

We choose the parameters of the model in such a way that
$\langle \psi^{\dagger} \psi \rangle \ll g_{6}^{-1}$; the
angular brackets denote averaging over the lattice. 
This means that in the initial state
the attractive interaction is much more important than the repulsive 
one. In this case, we expect that, in appropriate dimensionless units,
the time scale $t_{c}$ of the initial collapse 
of the gas into clumps depends only on the single remaining parameter 
of nonlinearity
\begin{equation}
\xi = \frac{\langle \psi^{\dagger} \psi \rangle}{2m \epsilon} \; ,
\label{xi}\end{equation}
where $\epsilon$ is the average kinetic energy per particle in the initial
state; $\xi$
is of order of the ratio of the initial potential energy of 
attraction to the initial kinetic energy. 
A similar parameter for an
atomic gas in a trap will be introduced below. We can write
\begin{equation}
t_{c}^{-1}= \epsilon F(\xi ) \; ,
\label{t0}
\end{equation}
where $F$ is some function obeying the condition $F(0)=0$ and $F(1) \sim 1$. 
The form of $F(\xi)$ at small $\xi$ is established below.
For the initial distribution (\ref{nk}), the initial parameter of 
nonlinearity is $\xi = k_{0} A / 12\pi^{3/2}$.
We choose units of time so that $2m=1$. We also use
units of length in which $k_{0}=2\pi$, i.e. measure lengths in 
units of the particles' typical initial de Broglie wavelength. Except where 
stated otherwise, we consider the 
case of moderate nonlinearity $A=5$, which corresponds to
$\xi= 0.47$. We use $g_{6}^{-1} = 3600$.

The results below are from integrations on a $64^{3}$ cubic lattice  
with side $L=2.25$ (in the above length units) and periodic boundary
conditions. The state of the system
was updated via a second-order in time algorithm based on the 
Crank-Nicholson method for the diffusion equation. The algorithm 
conserves the number of particles exactly. Energy non-conservation
was below 2\% for the entire integration time.

Fig. 1 shows two snapshots of
the field, at times $t=0.1$ and $t=20$. Dots have been placed on all lattice 
sites at which $|\psi|> 30$ (the mean-square value of $|\psi|$ is 5.3).
Drops of dew are clearly seen. 

\begin{figure}
\leavevmode\epsfysize=3.5cm \epsfbox{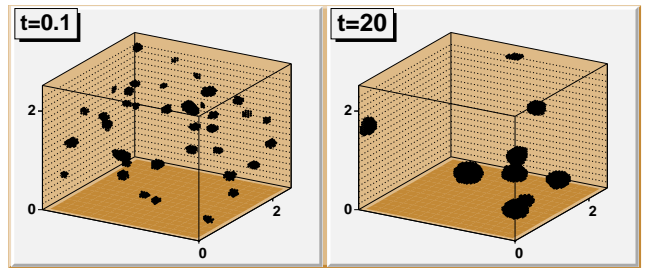}
\caption{Drops of dew at different moments of time.}
\label{fig:drops}
\end{figure}

A movie of the evolution of this
picture from $t=0.1$ to $t=20$ shows that the drops of dew
move around, gradually slowing down, and occasionally coalesce.
The overall growth of the number of sites with $|\psi|> 30$ continues even at
$t=20$, the latest time in our computation, but at that time it is already 
quite slow. 
If we define that grid points with $|\psi| \leq 8$ belong to the gas,
and correspondingly grid points with $|\psi| > 8$ belong to the dew 
($|\psi| = 8$ is approximately the boundary between the gas and the dew
at $t=20$, see Fig.~\ref{fig:PDF} below),
we find that around 15\% of all particles are in the gas, and around 85\% 
had condensed in the dew by the time $t=20$.
These fractions, however, may be altered when gravity is included.

\begin{figure}
\leavevmode\epsfysize=5.5cm \epsfbox{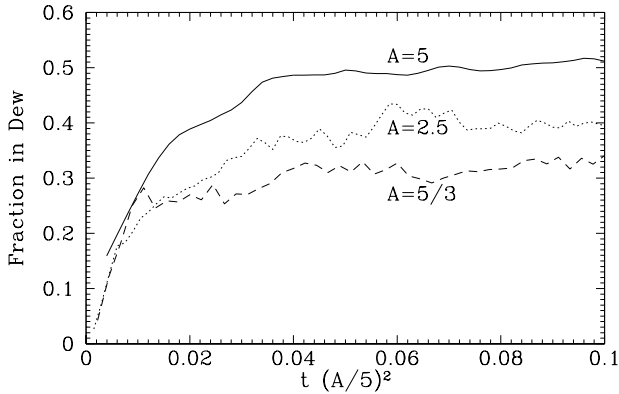}
\caption{Condensation process.}
\label{fig:cond}
\end{figure}

Fig. \ref{fig:cond} 
shows initial stages of the condensation process: we plot
the fraction of particles that are in the dew, as a function of time. 
We observe two 
distinct stages: a rapid collapse followed by a slower chaotic evolution. 
Because $\xi \sim 1$ at $A=5$, we estimate
the time of collapse $t_c$ from (\ref{t0}) as $t_c \sim k_0^{-2}$. For 
$k_0=2\pi$ this gives $t_c \sim 0.025$, in good agreement with the data of 
Fig.~\ref{fig:cond}. In the regime of weak nonlinearity, $\xi \ll 1$,
we expect the collapse to be due to two-particle collisions, in which case
$F(\xi ) \propto \xi^2$. (With this form of $F(\xi)$, the estimate 
(\ref{t0}) for the time of the collapse coincides with the estimate
of the condensation time of Ref. \cite{it91}, which can also be obtained
from a solution to the Boltzmann equation \cite{ST}).
The time of the collapse then has to scale as $A^2$ when we decrease $A$ 
and keep all other parameters
fixed. Results of integrations with different values of $\xi < 1$ confirm
this, see Fig. \ref{fig:cond}.

For an atomic gas confined in a trap, at some temperature $T$,
one can introduce initial parameter of nonlinearity $\xi_T$:
\begin{equation}
\xi_T = \frac{4\pi\hbar^2 |a| n}{m T} \; ,
\label{xiT}
\end{equation}
where $n$ is a typical gas density, and $a$ is the
scattering length, which in the present case is negative. 
In (\ref{xiT}), $k_B = 1$,
but we have restored $\hbar$. Let us use for estimates
$n = (m T_0 / 3.31 \hbar^2)^{3/2}$ and $T=T_0$, where $T_0$ is the
temperature of BEC of an ideal monoatomic gas in a given trap
and with a given number of particles.
Bradley {\it et al.} \cite{exp2} quote $T_0=300$ nK and
$a=-27.3 a_0$ for their experiment with trapped $^7$Li ($a_0$ is the
Bohr radius); using these values we obtain $\xi_T = 0.006$.
Estimating the rate of the collapse
as $\hbar t_c^{-1} \sim T_0 \xi_T^2$, we find $t_c\sim 1$ s.
We thus expect that quantum dew can be observed in laboratory in traps of
a sufficiently large size.

The onset of the slower chaotic evolution indicates that a chemical 
quasiequilibrium 
between the dew and the gas has been reached, i.e. the processes of 
evaporation of particles from the existing dew drops and condensation 
back onto them are approximately (but not exactly) balanced.
This interpretation is supported by the following test. The probability 
distribution of the absolute value of the field over lattice sites shows two
distinct peaks: one at large $|\psi|$, corresponding to the dew drops,
and another at small $|\psi|$, corresponding to the gas of particles,
see Fig. 3. 
If at some instant we
remove the gas, i.e. set $\psi=0$ at all sites where we had $|\psi|<10$,
and then continue the evolution, the gas reappears, while the number of 
sites occupied by the dew decreases down to another slowly evolving value. 
Apparently, the dew partially evaporates, so as to restore the chemical 
quasiequilibrium.

\begin{figure}
\leavevmode\epsfysize=5.5cm \epsfbox{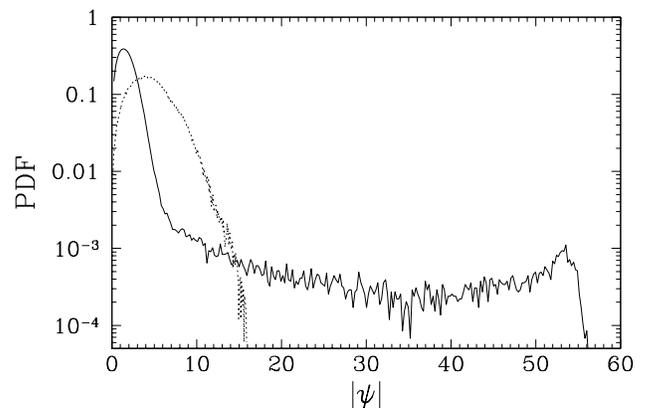}
\caption{Probability distribution function
of the field magnitude at $t=20$
(the solid line). Initial probability distribution is shown by the 
dotted line.}
\label{fig:PDF}
\end{figure}

\begin{figure}
\leavevmode\epsfysize=8.2cm \epsfbox{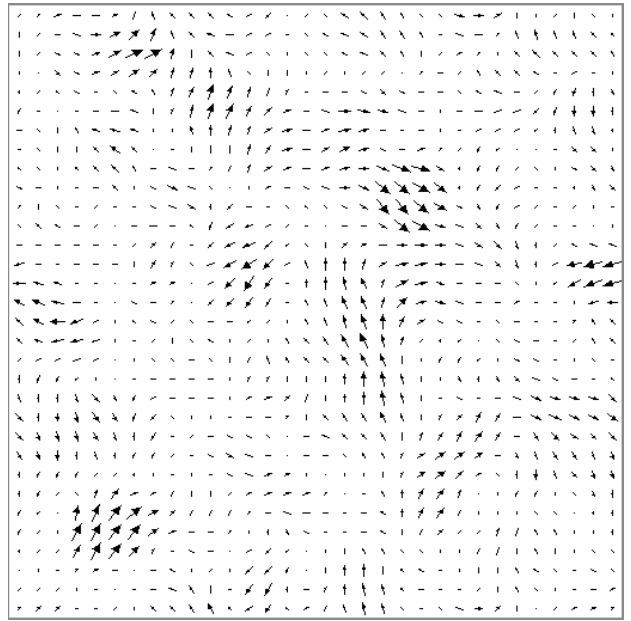}
\caption{Field distribution in a spatial slice at $t=0.2$.}
\label{fig:slice}
\end{figure}

Finally, Fig. 4 illustrates the coherent nature of the dew. It shows the 
field $\psi$ at $t=0.2$ at a section of our integration cube parallel to 
the $x$--$y$ plane. 
For visual clarity only even-even numbered sites are
included. The length of an arrow represents $|\psi|$, and the angle clockwise 
from 12 noon represents $\arg\psi$. We see that the sites occupied by the dew 
(i.e. having large $|\psi|$) are in drops, and each such drop is 
coherent---the arrows point approximately in the same direction.
The direction of arrows in each drop rotates with time, just as in the
homogeneous case \cite{Bogoliubov}, but these directions are different
for different drops. Similar slices at later times show that the clumping 
becomes more pronounced, the dew is still coherent, while the remaining gas 
(occupying sites with small $|\psi|$) is incoherent.

\begin{figure}
\leavevmode\epsfysize=5.5cm \epsfbox{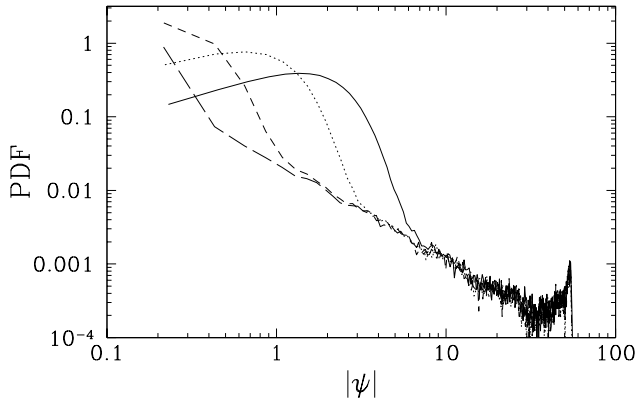}
\caption{P.d.f of the field's magnitude after truncating the gas and letting
it reequilibrate. Solid line: no truncation; dotted line: after truncation at 
$\psi = 10$; dashed line: after further truncation at $\psi = 3$; 
long dashed: after further truncation at $\psi = 1$.}
\label{fig:PDF_tr}
\end{figure}

Eq. (\ref{eqm}) has stable nontopological solitons
of the form
\begin{equation}
\psi(\r, t) = \chi(\r) \exp(i\omega t) 
\label{nontop}
\end{equation}
(nonrelativistic analogs of Q-balls \cite{qballs}). As we continue to 
truncate the gas, i.e. to remove
particles from sites with progressively smaller $|\psi|$, and to evolve the
system between these truncations, we expect to eventually reach a state in 
which
solitons float in vacuum (or gas of a very small density). Changes in the
probability distribution function
(p.d.f.) of $|\psi|$ resulting from this procedure 
are shown in Fig. \ref{fig:PDF_tr}. We interpret the limiting form to which the 
p.d.f. converges in the middle range of $|\psi|$ as corresponding to the wall 
profile of nontopological
solitons. Computation of $\psi$ at the center of a soliton in the thin-wall
approximation gives $|\psi|_c = (3/4g_6)^{1/2} \approx 52$, in good agreement 
with the position of the peak in the p.d.f. at large $|\psi|$. 
Like nontopological solitons produced in a decay of an unstable homogeneous
condensate \cite{KS}, quantum dew may work as cold dark matter.

To summarize, our main results are: (i) a numerical proof that the 
clumps of matter formed in a nonequilibrium gas with an attractive 
interaction are coherent drops of Bose-Einstein condensate; (ii) 
evidence that the rapid collapse of particles into drops of this
quantum dew is followed by a slower evolution, during which the dew
is in approximate chemical equilibrium with the surrounding gas;
(iii) evidence that at weak nonlinearity the rate of the initial collapse  
is consistent with being determined by two-particle collisions.

We thank A. Kusenko and M. Shaposhnikov for useful discussions.
This work was supported in part by DOE grant DE-FG02-91ER40681 (Task B)
and NSF grant PHY-9501458. S.K. thanks ITP, Santa Barbara for hospitality
during completion of this work.

\end{document}